# Hadoop Scheduling Base On Data Locality

*Jiang Bo. Wu Jiaying. Shi Xiuyu. Huang Ruhuan*
*Institute of Computer and Control*
*University of Chinese Academy of Science*
*{Jiangbo14, WuJiaying14, Shixiuyu14, Huangruhuan14}@mails.ucas.ac.cn*

**Abstract**: In hadoop, the job scheduling is an independent module, users can design their own job scheduler based on their actual application requirements, thereby meet their specific business needs. Currently, hadoop has three schedulers: FIFO, computing capacity scheduling and fair scheduling policy, all of them are take task allocation strategy that considerate data locality simply. They neither support data locality well nor fully apply to all cases of jobs scheduling. In this paper, we took the concept of resources-prefetch into consideration, and proposed a job scheduling algorithm based on data locality. By estimate the remaining time to complete a task, compared with the time to transfer a resources block, to preselect candidate nodes for task allocation. Then we preselect a non-local map tasks from the unfinished job queue as resources-prefetch tasks. Getting information of resources blocks of preselected map task, select a nearest resources blocks from the candidate node and transferred to local through network. Thus we would ensure data locality good enough. Eventually, we design a experiment and proved resources-prefetch method can guarantee good job data locality and reduce the time to complete the job to a certain extent.

## 0. Introduction

With the popularizing of Internet technology, whether it is business or personal generated data are in the rapid growth. Researches aim at how to effective and efficient mining useful knowledge from big-data to satisfied different business requirements had made a lot of achievements. The advent of the era of big-data, are making hadoop and Map-Reduce processing framework becoming increasingly popular, many companies and researchers are keen to study hadoop to meet their specific business needs. As one of the core technologies of hadoop, Map-Reduce job processing framework and job scheduling algorithm play a vital role in the overall performance of hadoop. In the dynamic task scheduling and resources allocation policies of hadoop, the input data will be cut into several pieces to storage on each node, and each node keep three copies in default. How to ensure that the data blocks needed is just located in the compute node within different tasks of a job during operation, and improve the utilization of system resources and efficiency of job operation, namely how to ensure good data locality, has become a hot issue in recent years.

This paper analyzed three commonly used hadoop job scheduling algorithms, FIFO, computing capacity scheduling, fair scheduling. Then introduced the necessity of data locality for hadoop job scheduling, and compare three types of commonly used scheduling algorithms in consideration of data locality. We proposed a data locality job scheduling policy based on resources-prefetch. We took full consideration of data locality of job, and achieved good data locality based on preselected nodes, prefetched tasks and resources. Finally, our experiment proves the feasibility of the proposed strategy and to assess the scheduling policy for hadoop performance improvements.

1. **Hadoop clustering job scheduling algorithm**

For the massive network transmission cost problem, hadoop takes mobile computing, not the design idea of mobile data. The job is divided into some map-tasks and reduce-tasks in the hadoop platform. Job scheduling refers to select suitable tasks which are suitable for jobs. Then they are distributed to appropriate computing nodes, Task-Tracker, and handled. Hadoop job scheduling frame has three stage scheduling framework. They are queue stage, job stage and task stage. At present, almost all hadoop job scheduling take the strategy.

FIFO scheduling algorithm is based on the order that a job is submitted to the job queue. The first in is on the head, the last in is on the end. It always starts the next job from the head when the current job is done. FIFO is simple, easy and cost little. Besides, it also makes it easy to distribute a job for Job-tracker. But it also has obvious defects. It doesn't take the distinct and urgency among jobs into account. Because of the strategy FIFO takes, first in first out, causes little jobs, behind large jobs, need to wait a long time and can't share the resources. It has a bad effect on the system's function. Moreover, FIFO job scheduling algorithm only considers the order job submitted, not the locality of resources that a job needed.

Fair scheduling strategy can make the job get equal right of sharing resources. Facebook brings it up. The strategy not only satisfies the different users' tasks in computing time, storage space, data flow and response time, but also uses the MapReduce framework to respond to a variety of parallel implementation of the job, and makes the user a good experience. When there is only one activity running, it will enjoy the whole resources alone. But when other jobs are submitted, Task-Tracker is released and assigned to the new job to ensure that each job is obtained. Fair-Scheduler uses resources pool to organize the job, in general, it allocates a resources pool for each user. Then the resources is fairly distributed to each resources pool, and in order to provide fair sharing way, Fair-Scheduler allows to allocate guaranteed minimum share resources for resources pool to ensure that users and cluster obtain enough resources. The calculation capacity dispatch could ensure that each job queue evenly gets calculation resources as far as possible, and to different degrees, improves the high usage rate of computing resources. Meanwhile, it also guarantees the fairness that the job attains calculation resources. However, the job scheduling strategy for each job queue is simply FIFO, it also makes the calculation ability scheduling only applies to the application scene that huge number of users and the user requests, and fairly to get the application of the resources.

Thus, scheduler from the current hadoop takes the task allocating strategy which simply considering local data. It can't support data's locality well, and the use of the scene is single, and is not fully applicable to all job scheduling occasions.

## 2. Data locality of Hadoop job

Data locality is a measure of task data localization. Based on the node position of input data of task, tasks can be divided into the node locality, rack locality and remote tasks.

Delay improved data locality greatly by setting a certain waiting time for non-local tasks and suspending their scheduling. However, this does not apply to long-running jobs, likely to cause obstruction. Guo and his fellows [1] proposed an algorithm that can schedule multiple tasks simultaneously to ensure data locality In the case of different applications need to share the input data and studied how data locality affected the job response time. Similarly, Bezerra et al [2] also consider the case of sharing the same resources among different Map Reduce applications, divided jobs that need to deal with the same data block into a group and assigned to the node where the data blocks it needs to run , once only dispatch a mission to ensure better data locality. Both have shared in the case of job input data good data locality, but relatively limited and does not apply to all job scheduling.

Data block prefetching is an effective way to hide access data loss caused by a buffer delay. Xie et al used data prefetching technology to prefetch input data from disk into memory before scheduling, which saves execution time when the task read data, thereby reducing job response time. Seo et al proposed scheduling strategy HPMP based on resources-prefetching and pre-shuffling to resolve the "deficiencies" that data locality cannot be guaranteed when multiple users share cluster resources on HOD.

In summary, a lot of present studies have improved data locality in order to improve hadoop job scheduling. This paper introduced the concept of data prefetching, guaranteed data locality of task effectively, and reduced the execution time of job.

## 3. Hadoop scheduling based on data locality

In this part, we proposed a job scheduling method based on data-locality mainly includes three parts, which are node-preselecting, task-preselecting and resources prefetching. Procedure of our algorithm have shown in Figure-1. The main idea is estimate the current remaining task execution time and then compared with the data block transmission time, and  the nodes that most likely to release the computing resources as preselected nodes. Then select a map task from job queue. We will select the task directly if local tasks existence, otherwise selecting the non-local task as the preselected task. Finally, according to the selected map tasks analysis of its resources block of the node, select the resources nodes that nearest to the preselected nodes and make resources block prefetched to local node through the network transmission. Thus ensure there is a local resources data copy when the task is assigned to the nodes.

**1) Node-preselect**

Firstly selecting the under prefetching computing nodes, mainly includes estimating the remaining time of the task, the time of estimating transmission between nodes and comparing both to choose which one is better. Using the rate of task executes to speculate the remaining time of the task, and each time choose a node with the smallest remaining time. For accomplishing the prefetching before the end of current task, we need to compare each $T_{left}$ and $T_{perblock}$ on each map task. These two values can be calculated according to the formula (1) and the formula (2).

$$T_{left} = \frac{1 - progress_t}{\Delta p_t} \quad (1)$$

Among them, $progress = \frac{finished}{total}$ represents task progress, which is estimated on the basis of resources block size and the block size read. Besides, $\Delta p_t = \frac{progress_t}{t}$ represents task progress rate, the $T$ represents the task being dispatched to the current execution time.

$$T_{perblock} = \frac{blockSize}{tranRate} \quad (2)$$

Getting it by estimates the cluster network bandwidth transmission rate. Set the candidate node set to *M*, if $T_{left} > T_{perblock}$, put the nodes into *M*, according to $T_{left} - T_{perblock}$, the small to large sort, selecting the smallest one as prefetching node *target_node*.

Because the internal mechanism of the hadoop has the number of failures on each node count to ensure that a point which goes through many failures are still assigned to the node. So we need a number of failed tasks if current node's failed tasks achieve the failed tasks that a job on the computing node for system. For timeliness, update a candidate node every heartbeat interval.

**2) Task-preselect**

The primary task is that choosing the waiting task on the basis of local data from those never run. In hadoop, *JobInProgress* object established and protected by Job-Tracker offers *failedMaps* and *nonRunningMapCache* of the current task. The strategy is follow:

I．To ensure the failed task first gets the computing resources, algorithm gets a task from *failedMaps* to schedule. If there is local task, then distribute to *target_node*, the pre selection node, and execute. Interrupt the algorithm execution and go on the next iteration. Or choosing a non local task and according to the local strategy and regarding remote tasks as waiting pre-selection task, *target_map*.

II. If the *failedMaps* is null, we would be better to select the task from the un-executed task queue. Like the first step, if there are local tasks in *nonRunningMapCache*, then directly distribute to *target_node*, the pre selection node, and execute. Interrupt the algorithm execution and go on the next iteration.

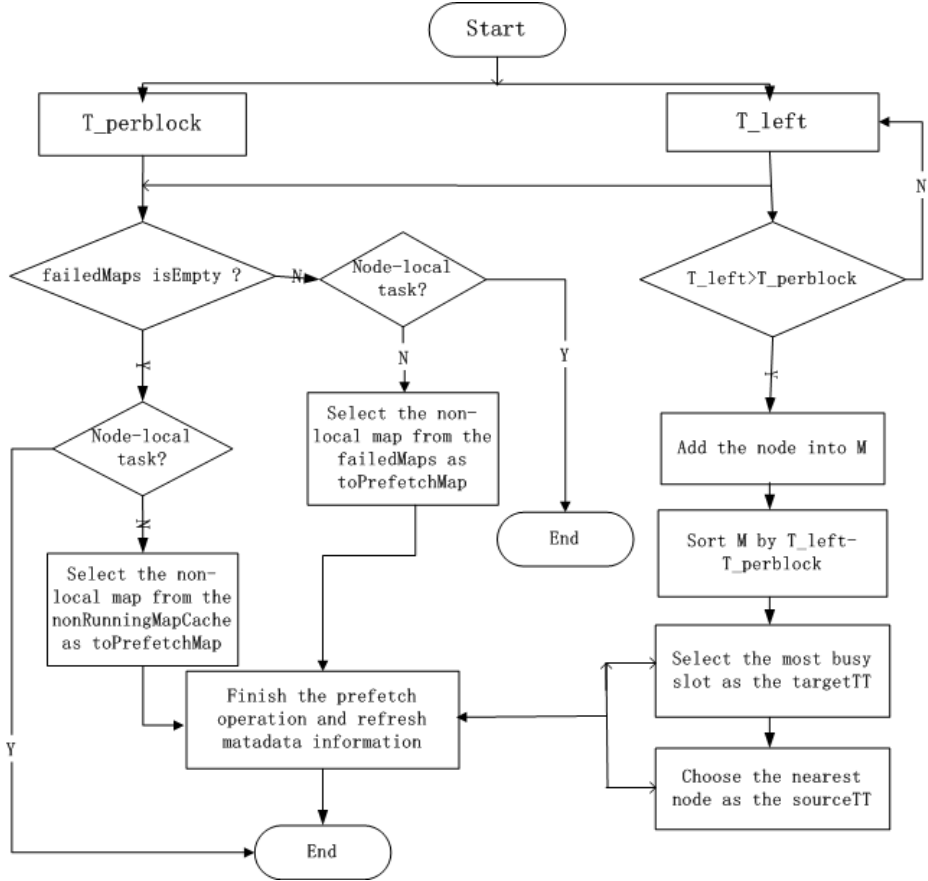

**Figure 1 algorithm flow chart**

### 3) Resources-prefetch

If the algorithm go to here, it represents the tasks waiting for prefetching on the (2) step aren't local tasks. Then we need to prefetch input data to candidate note target_node. It mainly includes determining the source node and updating metadata information. The prefetching process firstly reads metadata information according to the *TaskInProgress* which matches under prefetching task and confirms the candidate source node. Next, get the network topology information configured by current hadoop cluster. Determine the distance between the candidate source node and the destination node and then select the nearest point. Finally, transferring the input data to local, which is equal to increase data copies and update Job-Tracker's meta information file.

In order to avoid occupying too much resources, the algorithm only prefetch one node and one task each iteration. At the same time, when compute the distance between computing resources nodes and *target_node*. It doesn't take *Dijktra* algorithm to get optimum, but just compute the distance between different nodes by simple formula three.

## 4. Experiment result and analysis

The experiment test environment includes 4 PC, Hadoop0.20.2, one Job-Tracker nodes and three Task-Tracker nodes.

The experiment mainly configures and uses three improved schedulers and three schedulers hadoop had in the experiment cluster. The experimental scene is for three different users (*user0, user1, user2*) submitted- jobs at the same time. For the improved schedulers and computing capability schedulers, configuring a queue mapping each user. Using the commonly used *WordCount* as a test program, the situation each user submits is as shown in Table-1. The measurement of the scheduling algorithm is mainly local data locality and job response time. The former can observe the log of the accomplished job, and get it by the statistic proportion of local tasks in total tasks. And the latter can get it from the log of the accomplished job. Configuring and using 4 different schedulers the above mentioned in order in experimental hadoop cluster. Submit all jobs in the Table-1 in the same way and the same order, and observe the results of the operation, and get the following results by statistical analysis.

**Table 1 job related information**

| Job-name | User  | Data | Map | Reduce |
|----------|-------|------|-----|--------|
| Job0     | User0 | 512M | 14  | 1      |
| Job1     | User0 | 512M | 14  | 1      |
| Job2     | User0 | 512M | 14  | 1      |
| Job3     | User1 | 1G   | 16  | 1      |
| Job4     | User1 | 1G   | 16  | 1      |
| Job5     | User1 | 1G   | 16  | 1      |
| Job6     | User2 | 2G   | 32  | 1      |
| Job7     | User2 | 2G   | 32  | 1      |

As figure 1 shows, in the same experiment scene, the three job scheduling strategy are not quite different from the task data locality, are keeped at 40%-50%. This is because the three scheduling policies are simply compared to the task level scheduling stage to see whether the task is local. But by the node-preselecting, map task preselecting, resources-prefetch, and the three node local guarantee can greatly improve the task data's locality, and improve ranging from 15% to 20%, greatly improve resources utilization rate.

**Fig. 1 experimental results of local data**

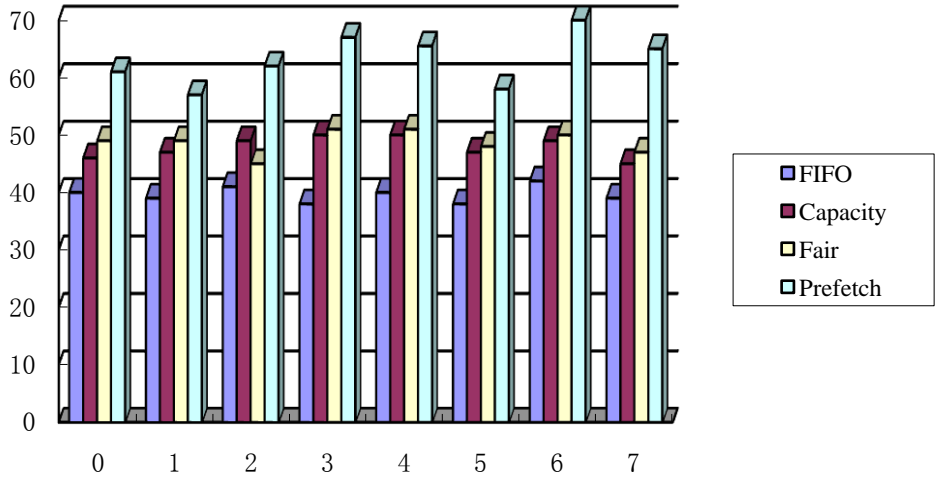

As shown in Figure 2, the job scheduling algorithm we proposed, its job response time and computing power scheduling algorithm maintain fairly. But resources-prefetch reduces the time that non local tasks wait for local nodes' appear. This also shows that the resources-prefetch not only improve the local data of map tasks, but also decrease the job response time.

**Figure 2 job response time**

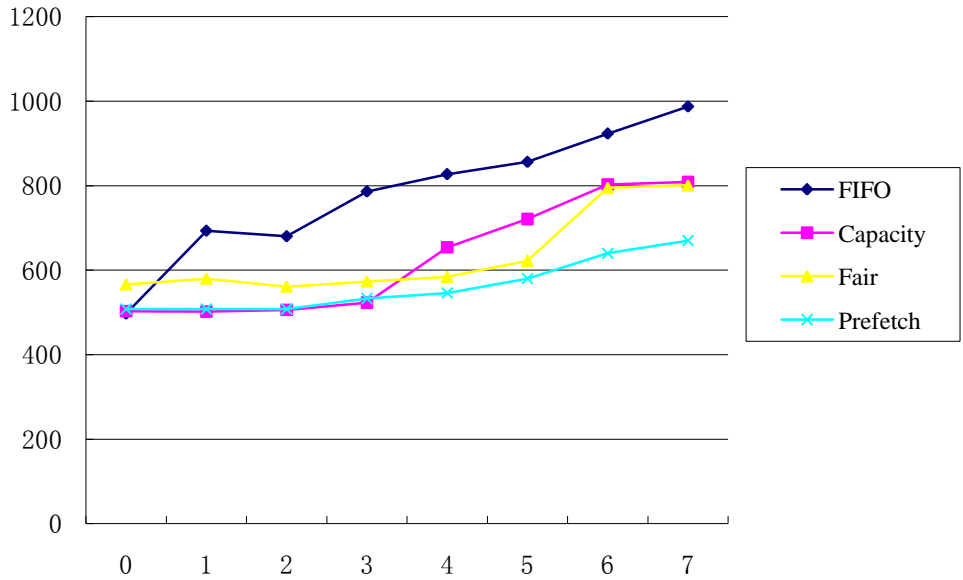

In conclusion, we have experimented on small hadoop clusters and compared with original job scheduling algorithm of hadoop. And it presents that the job scheduling algorithm proposed in this paper not only improves the data locality of the task, but also reduces the response time of the job. This also proves the algorithm's feasibility put forward in the paper.

## 5. Conclusion

We analyzed three types of commonly used hadoop job scheduling algorithm deficiency in terms of local data. Since the mobile computing design of hadoop cluster, we believe that when the map had to assign the task to a non-local node, time the node wait for locality or tasks will be wasted. Finally, not only it caused a lot of network overhead, and the efficiency of operations and resources utilization will be decreased. Therefore, we refer to the concept that prefetch data for the relief of data access latency. Before non-local map task be assigned, we prefetchs its resources to the local candidate nodes with certain network I/O and disk space overhead to make data locality good enough. By implement this idea in hadoop0.20.2, and finally experiment on small hadoop cluster. The results presents that our algorithm improves data locality of map tasks about 15% significantly, and decreases job response time to some extent. Thus we reach the goal that improvement on system resources utilization.